# Gefitinib-Induced Interface Engineering Enhances the Defect Formation Energy for Highly Efficient and Stable Perovskite Solar Cells

*Xianhu Wu,[1] Guanglei Cui,[1,]\* Jieyu Bi,[1] Gaojie Xia,[1] Zewen Zuo,[1] Min Gu[2]*

[1] *College of Physics and Electronic Information, Anhui Province Key Laboratory for Control and Applications of Optoelectronic Information Materials, Key Laboratory of Functional Molecular Solids, Anhui Normal University, Wuhu 241002, P. R. China.*

[2] *National Laboratory of Solid State Microstructures, Nanjing University, Nanjing 210093, P. R. China.*

\* Corresponding Author.

Electronic mail: wxh@ahnu.edu.cn (X. Wu); glcui@ahnu.edu.cn (G. Cui)

**ABSTRACT**

Poly(3,4-ethylenedioxythiophene):polystyrenesulfonate (PEDOT:PSS) has been widely used as a hole transport layer in perovskite solar cells (PSCs). However, the high interface defect density and energy level mismatch between PEDOT:PSS and perovskite can lead to significant open-circuit voltage loss. Additionally, the free PSS chains on the surface of PEDOT:PSS can absorb water molecules, promoting the degradation of perovskite at the PEDOT:PSS/perovskite interface. Here, gefitinib is used to modify the surface of PEDOT:PSS, removing a portion of the free PSS chains from the surface, reducing the PSS/PEDOT ratio, and enhancing the conductivity of PEDOT:PSS. Gefitinib has altered the energy level structure of PEDOT:PSS, facilitating hole transport at the interface. The Cl, F, and NH groups in gefitinib also passivated defects in the perovskite, reducing the defect density at the interface and significantly enhancing the stability of PSCs. This modification increased the open-circuit voltage from 1.077 to 1.110 V and the power conversion efficiency (PCE) from 17.01% to 19.63%. When gefitinib was used to modify the interface between $SnO_2$ and perovskite, the PCE of PSCs (ITO/$SnO_2$/perovskite/Spiro-OMETAD/Au) increased from 22.46% to 23.89%. This approach provides new perspectives and strategies for improving the efficiency and stability of PSCs.

# I. INTRODUCTION

Perovskite materials are widely used in photovoltaic devices, photodetectors, and light-emitting diodes due to their tunable bandgap, excellent optoelectronic properties, low exciton binding energy, and long exciton diffusion length. [1-3] In just over a decade, the power conversion efficiency (PCE) of perovskite solar cells (PSCs) has increased from 3.9% to over 26%. [4,5] Inverted PSCs, due to their advantages such as ultra-thinness, wear resistance, ease of low-temperature fabrication, and reduced hysteresis, exhibit significant commercial potential. In PSCs, the commonly used hole transport layers (HTL) include 2,2',7,7'-tetrakis(N,N-di-pmethoxyphenylamine)-9,9'-spirobifluorene (Spiro-OMeTAD), poly[bis(4-phenyl) (2,4,6-trimethylphenyl) amine] (PTAA), nickel oxide, and Poly(3,4-ethylenedioxythiophene): polystyrenesulfonate (PEDOT:PSS). Spiro-OMETAD requires the addition of Li salts and TBP to enhance the conductivity of the HTL, PTAA has poor surface wettability, which hinders the nucleation and crystallization of perovskite films on the surface, and nickel oxide has poor optical transmittance, which negatively affects the light absorption of the perovskite. [5-9] Compared to these materials, PEDOT:PSS has become a strong competitor for the HTL in inverted PSCs due to its low cost and high optical transmittance. PEDOT:PSS is now widely used in inverted PSCs, especially in tin-lead-based PSCs. However, the uncoordinated PSS molecules on the surface of PEDOT:PSS, when in direct contact with perovskite, can cause degradation of the perovskite at the interface, which is detrimental to the long-term stability of the perovskite. Furthermore, the interface between PEDOT:PSS and perovskite can form severe non-radiative

recombination centers, hindering charge transport at the interface. [10-12]

To address these issues, the widely used approaches currently involve doping PEDOT:PSS and modifying the interface between PEDOT:PSS and perovskite. While doping PEDOT:PSS can regulate the energy levels of PEDOT:PSS and improve the alignment with the perovskite, it also facilitates interfacial charge transfer. However, the surface of PEDOT:PSS still directly contacts the perovskite. Therefore, interface modification between PEDOT:PSS and perovskite improves the long-term stability of the perovskite. Compared to doping PEDOT:PSS, organic materials effectively avoid direct contact between PEDOT:PSS and perovskite. Extensive research has been conducted by scholars on interface modification with organic materials. For example, [2-(9H-carbazol-9-yl)ethyl]phosphonic acid was used to modify PEDOT:PSS, resulting in improved energy level alignment between PEDOT:PSS and the perovskite, enhanced perovskite layer compactness, and reduced non-radiative recombination at the PEDOT:PSS/perovskite interface.[13] 2-Chloroethyl phosphonic acid (CEPA) was employed to modify PEDOT:PSS. CEPA reduces the excess PSS on PEDOT:PSS, minimizes voids and impurity phases at the buried interface, and decreases degradation at the tin-based perovskite/PEDOT:PSS interface.[14] Polyethylene glycol-200 was utilized to improve the filling of PEDOT:PSS polymer chains, significantly reducing the interfacial charge transport resistance.[15] Hydrophobic fluorinated derivatives were employed to enhance the hydrophobicity and work function of PEDOT:PSS, alleviate the formation of charge recombination centers on the perovskite surface, and reduce energy loss at the interface.[16] Hydroquinone (HQ) was used to treat the surface of PEDOT:PSS, reducing the hole transport barrier at the interface and consequently lowering the interfacial resistance. The leakage current in PEDOT:PSS devices treated with HQ is significantly reduced, and the surface modification improves the interfacial contact.[17] Poly(triethyleneamine) was employed to modify the PEDOT:PSS/perovskite interface, significantly reducing the open-circuit voltage loss in tin-lead PSCs.[18] Polyethyleneimine ethoxylated was utilized to modify the perovskite/ PEDOT:PSS interface, neutralizing the PSS on the surface of PEDOT:PSS films, which not only reduces the acidic corrosion of the perovskite but also significantly improves the device efficiency.[19] The low thermal hysteresis of ortho-carborane was used to promote hole extraction at the perovskite/PEDOT:PSS interface and reduce charge recombination, thus enhancing the thermal stability of tin-lead perovskite.[20] Ammonium sulfide (AS) was employed to modify the buried interface of the perovskite/PEDOT:PSS, where AS optimizes perovskite crystallization by forming PbS nucleation sites and reacts with PSS to neutralize the acidity of PEDOT:PSS. This process thereby tunes the energy levels of PEDOT:PSS, promoting charge transport at the perovskite/PEDOT:PSS interface.[21] Pentylamine acetate (PenAAc) was used to reduce interface defects between PEDOT:PSS and Sn-Pb perovskite, enhancing the bonding between the two materials. Additionally, the PenAAc interlayer improves charge extraction and inhibits the oxidation of $Sn^{2+}$ to $Sn^{4+}$.[22]

In this article, the organic molecule gefitinib (Gftn) was used to modify the surface of PEDOT:PSS, which removed the free PSS chains on the surface, reduced the PSS-to-PEDOT ratio, and enhanced the conductivity of PEDOT:PSS. This modification facilitated carrier transport at the PEDOT:PSS/perovskite interface. Gefitinib contains F, Cl, and NH groups, which passivated defects in the perovskite, reducing the defect density at the interface and significantly improving the stability of PSCs. The results showed that the open-circuit voltage increased from 1.077 to 1.110 V and the PCE increased from 17.01% to 19.63%. When gefitinib was applied to modify the interface between $SnO_2$ and perovskite, the PCE of the PSCs (ITO/$SnO_2$/perovskite/Spiro-OMETAD/Au) increased from 22.46% to 23.89%. This method provides new perspectives and strategies for improving the PCE and stability of PSCs.

## II. RESULTS AND DISCUSSION

The preparation process of gefitinib interface modification in PSCs was demonstrated in **Figure 1**. The XPS spectra of PEDOT:PSS (Control) and gefitinib-modified PEDOT:PSS (3-Gftn) were first measured and are presented in **Figure S1**. The XPS spectra of Cl elements for Control and 3-Gftn are shown in **Figure 2a**, directly demonstrating the successful preparation of gefitinib on the surface of PEDOT:PSS. The XPS spectra of S elements for Control and 3-Gftn are shown in **Figure 2b**. It was estimated that the ratio of PSS chains to PEDOT decreased from 5.29 to 4.56. In PEDOT:PSS, PSS is the non-conductive part, and the reduction in this ratio was found to be beneficial for enhancing the conductivity of PEDOT:PSS. [23] The conductivities of Control and 3-Gftn are shown in **Figure 2c**. Compared to PEDOT:PSS, the conductivity of PEDOT:PSS modified with gefitinib was increased, which was attributed to the removal of some free, non-conductive PSS chains from the surface of

PEDOT:PSS after gefitinib modification. To investigate the effect of gefitinib on the optical properties of the PEDOT:PSS surface, the optical transmittance spectra of Control and 3-Gftn are shown in **Figure 2d**. 3-Gftn enhances the light transmittance of PEDOT:PSS in the region below 630 nm, which is beneficial for allowing more photons to enter the perovskite. There is a slight decrease in transmittance above 630 nm, but the reduction is minimal. Considering the perovskite bandgap and its absorption strength in this range, this will not significantly affect the light absorption of the perovskite. The optical absorption spectra of Control and 3-Gftn were shown in **Figure 2e**. Both absorption peaks in the figure originate from the PSS chains in PEDOT:PSS. The significantly reduced absorption peak of the PSS chains in 3-Gftn further confirms the removal of some free PSS chains from the PEDOT:PSS surface after gefitinib modification. To investigate the effect of gefitinib modification on the energy level structure of PEDOT:PSS, the ultraviolet photoelectron spectra (UPS) of PEDOT:PSS and 3-Gftn were measured and shown in **Figure S2**. The $E_{cut\text{-}off}$ values for PEDOT:PSS and 3-Gftn were found to be 16.34 and 16.14 eV, respectively. Using the formula $E_F = E_{cut-off} - 21.22\ eV$, the work functions of PEDOT:PSS and 3-Gftn were calculated to be -4.88 and -5.08 eV, respectively. The Fermi level edges ($E_{F,edge}$) for PEDOT:PSS and 3-Gftn were determined to be 0.21 and 0.18 eV, respectively. Using the formula $E_{VB} = E_F - E_{F,edge}$, the valence band maxima (VBM) of PEDOT:PSS and 3-Gftn were found to be -5.09 and -5.26 eV, respectively. Based on these parameters, the energy level structures of PEDOT:PSS and 3-Gftn were obtained and shown in **Figure 2f**.

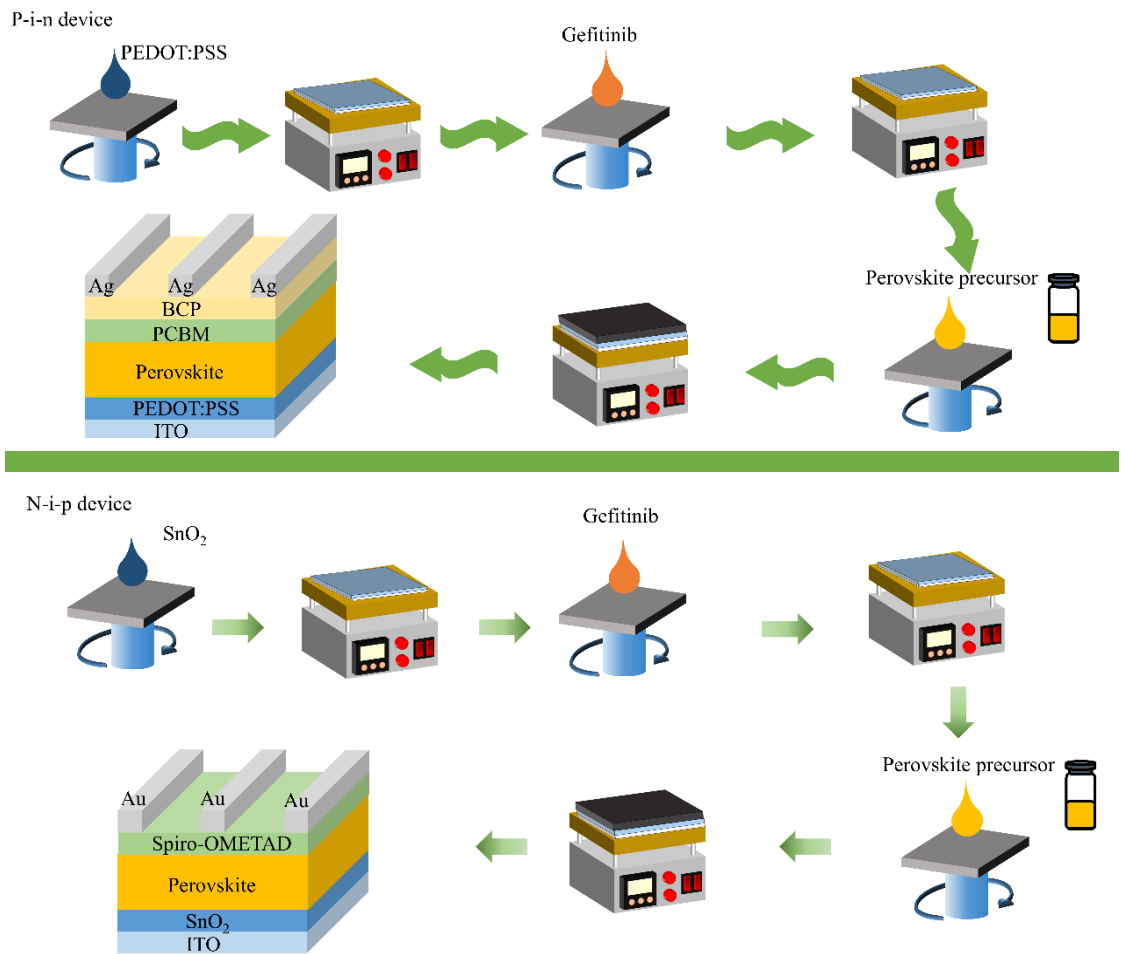

**Figure 1**. The Preparation process of gefitinib interface modification.

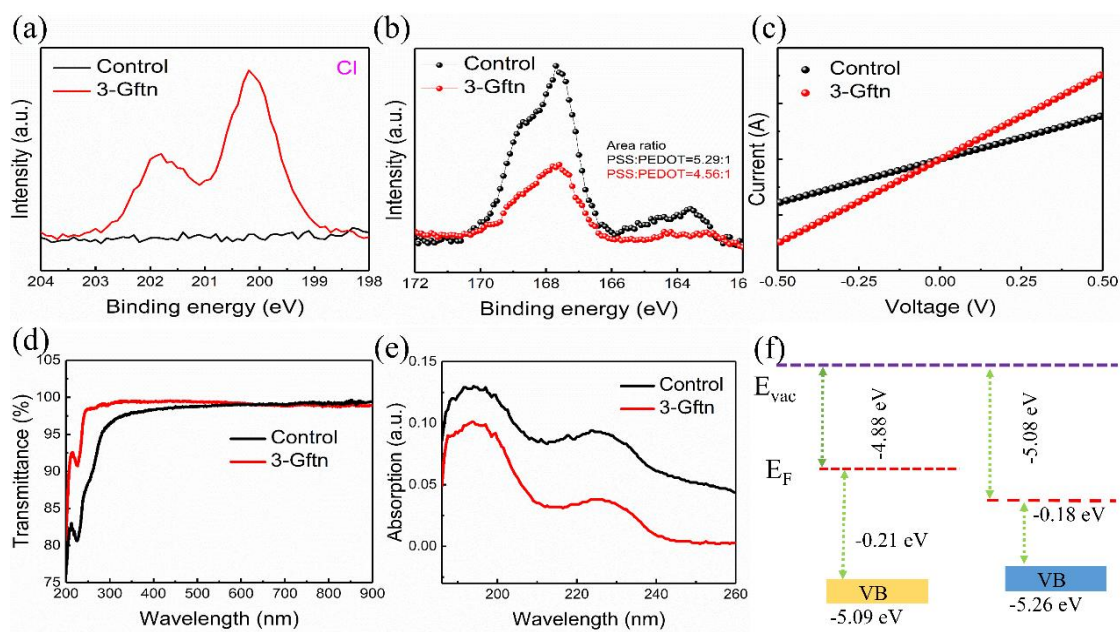

**Figure 2**. The XPS spectra of the (a) Cl element and (b) S element for Control and 3-

Gftn. (c) The conductivity of Control and 3-Gftn. (d) The optical transmission spectra of Control and 3-Gftn. (e) The optical absorption spectra of Control and 3-Gftn. (f) The energy level structure of Control and 3-Gftn.

Based on previous studies, the energy level diagram of the PSCs was determined and shown in **Figure S3**. It was observed that the VBM of 3-Gftn was closer to the VBM of the perovskite, which was more favorable for hole transport at the PEDOT:PSS/perovskite interface. The photoluminescence (PL) spectra of perovskite films prepared with PEDOT:PSS and 3-Gftn were shown in **Figure 3a**. The PL quenching in the perovskite film based on 3-Gftn occurs more quickly than that in the perovskite film based on PEDOT:PSS, directly indicating that hole transport at the PEDOT:PSS/3-Gftn/perovskite interface is faster. The electrostatic potential (ESP) of gefitinib was displayed in **Figure 3b**, where regions with higher negative electrostatic potential, including the NH group, F, and Cl, were shown. These regions were found to be beneficial for passivating uncoordinated defects in the perovskite.[24-27] To further investigate the interaction between gefitinib and perovskite, the XPS spectra of the Pb 4f peaks for the Control and gefitinib-modified perovskite were displayed in **Figure S4**. In the Control, the binding energies of Pb 4f are 143.30 and 138.49 eV, while in the gefitinib-modified perovskite, the binding energies of the Pb 4f peaks are 143.10 and 138.29 eV. After modification with gefitinib, the binding energies of the Pb 4f peaks shift 0.20 eV towards lower binding energy, suggesting that an interaction exists between gefitinib and the perovskite.[28-30] To investigate the effect of gefitinib on the morphology of perovskite, the scanning electron microscope (SEM) images of the

Control and 3-Gftn-based perovskites were measured and shown in **Figures 3c and 3d**. Compared to the Control, the surface of the 3-Gftn perovskite film shows fewer pinholes, which may be attributed to the fact that gefitinib altered the growth environment of the perovskite, thereby promoting perovskite crystallization. The grain size distribution within the range of the statistical ranget was displayed in **Figure S5**, and it is found that the average grain size of the perovskite increased from 220.20 to 299.70 nm. Larger grains mean fewer grain boundaries and defects, which is more favorable for the transport of charge. To investigate the effect of gefitinib-modified PEDOT:PSS on the crystallinity of perovskite, the x-ray diffraction (XRD) patterns of the Control and 3-Gftn perovskites were measured and shown in **Figure 3e**. The intensities of the (110), (220) and (310) diffraction peaks in the perovskite film based on 3-Gftn are significantly higher than those in the Control, indicating that gefitinib promotes the crystallization of perovskite. To explore the effect of gefitinib-modified PEDOT:PSS on the optical absorption of the perovskite, the optical absorption spectra of the Control and 3-Gftn perovskites were measured and displayed in **Figure 3f**. The optical absorption of the perovskite film based on 3-Gftn is significantly higher than that of the Control, which is beneficial for generating more photogenerated carriers and increasing the photocurrent of the device. The corresponding tauc plots based on the optical absorption spectra of the perovskite were displayed in **Figure S6**. After modification with gefitinib, the bandgap of the perovskite increases slightly from 1.613 to 1.615 eV.

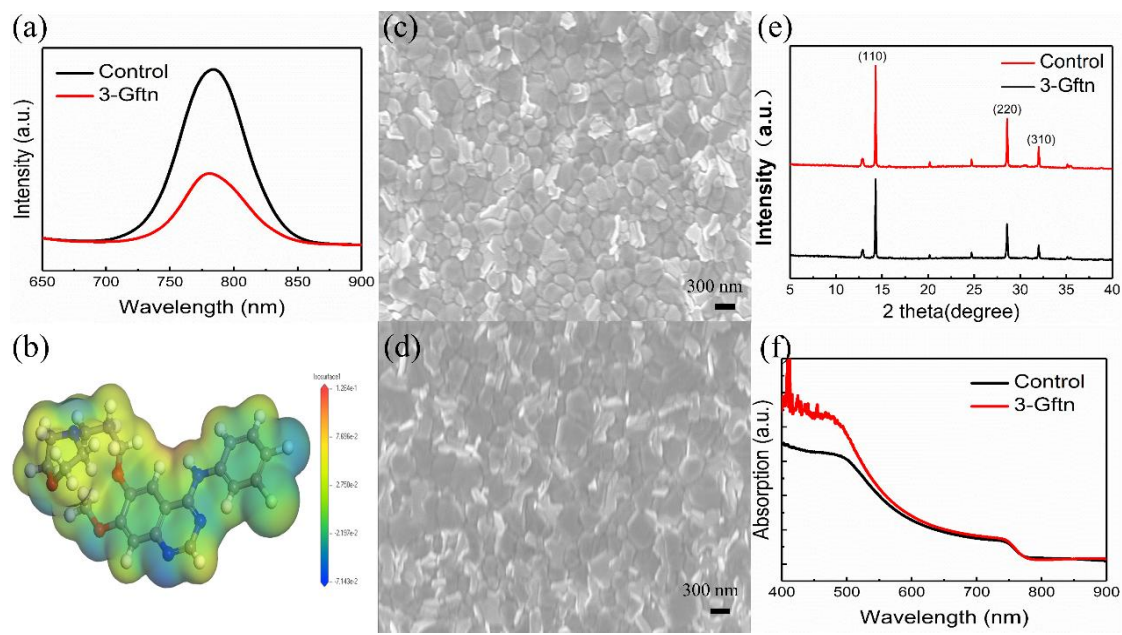

**Figure 3**. (a) Steady-state PL of perovskite films on Control and 3-Gftn. (b) ESP distribution of Gftn molecule. Top-SEM of perovskite films on (c) Control and (d) 3-Gftn. (e) XRD of perovskite films on Control and 3-Gftn. (f) Optical absorption spectra of perovskite films on Control and 3-Gftn.

To investigate the effect of gefitinib-modified PEDOT:PSS on the defects in perovskite, the space-charge-limited current (SCLC) of the devices based on Control and 3-Gftn was measured and displayed in **Figure S7**. The space charge limited voltage ($V_{TFL}$) of the device decreased from 0.787 to 0.701 V. The defect density of the devices based on Control and 3-Gftn was calculated using the formula $N_t = (2\varepsilon_0 \varepsilon V_{TFL})/eL^2$, where L is the perovskite film thickness (450 nm), $\varepsilon$ is the relative permittivity of perovskite, $\varepsilon_0$ is the vacuum permittivity, and e is the elementary charge. The calculated defect densities of the devices based on Control and 3-Gftn are $1.33 \times 10^{16}$ and $1.18 \times 10^{16}$ cm$^{-3}$, respectively (**Table S1**). The significant reduction in defect density is attributed to the passivation of uncoordinated defects at the PEDOT:PSS/perovskite interface after modification with gefitinib, which reduces non-radiative recombination.

To investigate the effect of gefitinib-modified PEDOT:PSS on device performance, the champion J-V curves of the devices based on Control and 3-Gftn were displayed in **Figure 4a**. The open-circuit voltage ($V_{oc}$), short-circuit current density ($J_{sc}$), fill factor (FF), and PCE of the Control-based devices are 1.077 V, 21.14 mA cm$^{-2}$, 74.73%, and 17.01%, respectively. The photovoltaic performance of the 3-Gftn-based devices shows a significant improvement, with $V_{oc}$, $J_{sc}$, FF and PCE reaching 1.110 V, 22.39 mA cm$^{-2}$, 78.96%, and 19.63%, respectively. The substantial enhancement in device performance is attributed to the better energy level alignment between the 3-Gftn-modified PEDOT:PSS and perovskite. The photovoltaic parameters of 30 devices based on the Control and different concentrations of gefitinib are presented in **Table S2**. As shown in **Figure 4b**, the integrated $J_{sc}$ of the EQE increased from 20.55 to 21.64 mA cm$^{-2}$, confirming the change in $J_{sc}$ observed in **Figure 4a**. The J-V curves of the devices based on Control and 3-Gftn under different scanning directions were displayed in **Figure 4c**. By estimation, the hysteresis coefficient of the devices based on Control and 3-Gftn decreases from 0.03 to 0.02, suggesting that gefitinib modification of PEDOT:PSS reduces the hysteresis of the devices. The dark J-V curves of the devices based on Control and 3-Gftn was displayed in **Figure 4d**. Notably, the 3-Gftn-based devices exhibit lower current compared to the Control, indicating that the leakage current is reduced in the 3-Gftn-based devices. Furthermore, when gefitinib was introduced at the SnO$_2$/perovskite interface, the PCE of the gefitinib-modified PSCs (ITO/SnO$_2$/perovskite/Spiro-OMETAD/Au) increased from 22.46% to 23.89% (**Figure**

**4e)**. The stability of PSCs in ambient air is also crucial. Therefore, long-term stability tests of unencapsulated devices were conducted under conditions of approximately 35±5% humidity. As shown in **Figure 4f**, after about 900 hours, the Control-based devices only retained 11.42% of their initial PCE, whereas the 3-Gftn-based devices maintained 52.31% of their initial PCE. The enhanced stability of the 3-Gftn-based devices is likely due to the reduced surface defects, which lead to better crystallinity of the perovskite film and consequently improve stability.

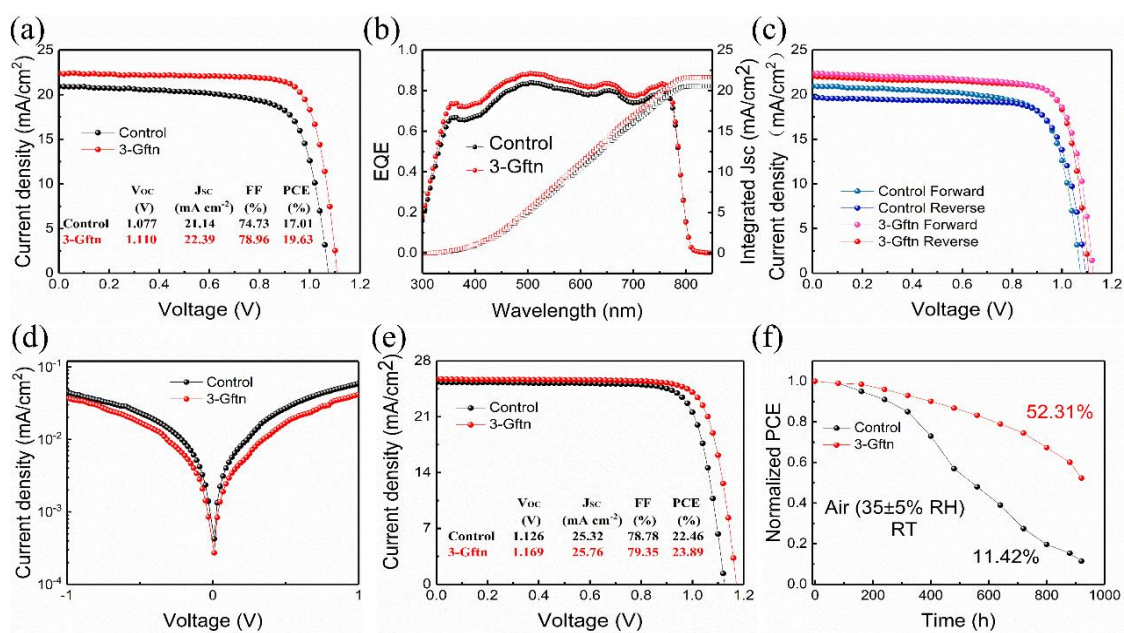

Figure 4. (a) The champion J-V curves and (b) EQE of p-i-n devices based on Control and 3-Gftn. (c) The J-V curves of devices based on Control and 3-Gftn in different scanning directions. (d) Dark J-V curves of devices based on Control and 3-Gftn. (e) The champion J-V curves and of n-i-p devices based on Control and 3-Gftn. (f) Stability in air ambient with RH of 35 ± 5% without encapsulation of p-i-n devices based on Control and 3-Gftn.

## III. CONCLUSIONS

In summary, this article employs gefitinib-modified PEDOT:PSS surfaces. After modification with gefitinib, the ratio of PSS chains to PEDOT chains on the surface of PEDOT:PSS decreased, resulting in an improvement in its conductivity. Additionally, this modification brings the valence band maximum of PEDOT:PSS closer to that of the perovskite, facilitating hole transport at the PEDOT:PSS/perovskite interface. As a result, the $V_{oc}$ of the device increases from 1.077 to 1.110 V, and the PCE significantly rises from 17.01% to 19.63%. This method provides a novel approach for organic molecule modification of interfaces to improve the PCE and stability of inverted PSCs.

## SUPPLEMENTARY MATERIAL

See the supplementary material for the experimental section (Materials, Device fabrication, Characterization). The XPS and UPS of PEDOT:PSS. The energy level structure of the device. The XPS and tauc plot of perovskite. The SCLC of device. The average photovoltaic performance of 30 devices based on Control and 3-Gftn.

## ACKNOWLEDGMENTS

This work was supported by the National Natural Science Foundation of China through Grants 21373011 and 12264060, Anhui Provincial Natural Science Foundation (2108085MA24).

## AUTHOR DECLARATIONS
### Conflict of Interest

The authors have no conflicts to disclose.

### Author Contributions

**Xianhu Wu**: Data curation (lead); Formal analysis (equal); Investigation (lead); Methodology (equal); Writing – original draft (equal); Writing – review & editing (equal). **Guanglei Cui**: Conceptualization (equal); Funding acquisition (equal); Methodology (equal); Supervision (equal); Writing – original draft (equal); Writing – review & editing (equal). **Jieyu Bi**: Conceptualization (equal); Methodology (lead); Supervision (equal); Writing – original draft (equal). **Nian Liu**: Investigation (supporting). **Gaojie Xia**: Data curation (supporting). **Jilong Sun**: Investigation

(supporting). **Luyang Liu**: Data curation (supporting). **Ping Li**: Data curation (supporting). **Zewen Zuo**: Methodology (equal). **Min Gu**: Investigation (supporting).

‡These authors contributed equally.

**Corresponding Author**

https://orcid.org/0009-0006-5831-993X **E-mail:** wxh@ahnu.edu.cn

https://orcid.org/0000-0002-3993-4543 **E-mail:** glcui@ahnu.edu.cn

**DATA AVAILABILITY**

The data that support the findings of this study are available within the article and its supplementary material.